\documentstyle[eqsecnum,epsf,prd,aps,preprint,tighten]{revtex}
\newcommand{\beq}{\begin{equation}}
\newcommand{\eeq}{\end{equation}}
\newcommand{\bea}{\begin{eqnarray}}
\newcommand{\eea}{\end{eqnarray}}

 \def\t{\tilde}
\def\m{\mu}
\def\s{\sigma}

\def\l{\lambda}

\preprint{\hfill\parbox{4cm}{SNUTP 00-002\\
                             UT-KOMABA 00-01\\}}

\begin{document}
\draft \title{Exact Zeros of the Partition Function for a Continuum System with Double Gaussian Peaks}
\author{Julian Lee\footnote{jul@hep1.c.u-tokyo.ac.jp}}
\address{Institute of Physics, University of Tokyo \\                
 Komaba, Meguro-ku, Tokyo 153, Japan}
\author{Koo-Chul Lee\footnote{kclee@phya.snu.ac.kr}}
\address{Department of Physics and Center for Theoretical Physics \\      Seoul National University,               
 Seoul, 151-742, Korea}
\date{\today}
\maketitle 
\begin{abstract}
  We calculate the exact  zeros of the partition function for a continuum system where the probability distribution for the order parameter is given by two asymmetric Gaussian peaks. When the positions of the two peaks coincide, the two separate loci of zeros which used to give first-order transition touch each other, with density of zeros vanishing at the contact point on the positive real axis. Instead of the second-order transition of Ehrenfast classification as one might naively expect, one finds a critical behavior in this limit.   
\end{abstract}

\setcounter{footnote}{0}
\narrowtext 

\section{Introduction}
It has been a central theme since the discovery of
statistical mechanics to  understand how the analytic partition
function for a finite-size system acquires a singularity in the thermodynamic limit if the system undergoes a
phase transition\cite{yang72}. The Lee-Yang theory \cite{ly}
has partly furnished the answer to this quest.   They proposed a scenario where the zeros of the partition function form a line and cut across the real axis. They showed that the discontinuity in the 1st order derivative of partition function is proportional to the angular density of zeros, using an analogy with the two dimensional electrostatics. Then they proved this scenario for Ising-like discrete systems under very general conditions. They could show that the zeros were distributed on a unit circle in this case. 
 
There have been  many attempts to generalize the ``Lee-Yang circle theorem" ever since.  Fisher
\cite{fisher65} initiated a study of zeros of the partition function in the complex temperature plane and  extensive studies of zeros of the 
partition function in complex temperature plane  followed  
\onlinecite{ono68,pearson82,abe67,itzykson83,privman87,bhanot87,kcl93,kcl94,kcl99}. In these works authors considered  continuous phase transitions or critical points.  

The conceptual basis of the Lee-Yang circle theorem was finally clarified in ref.\cite{kcl}  by considering the first-order transition of a system with more general continuous degrees of freedom, with a doubly peaked probability distribution for the order parameter. Since the Ising-like models considered by Lee and Yang would be described by two symmetric Gaussian peaks in the thermodynamic limit, this result provides a simple conceptual basis for Lee-Yang unit circle theorem. Furthermore it is a generalization since general asymmetric configurations were considered, whose zeros form a curve which is not a unit circle in general.

\begin{figure}
\epsfxsize=8.00cm
\epsfysize=6.00cm
\vskip0.1cm
\centerline{
\epsfbox{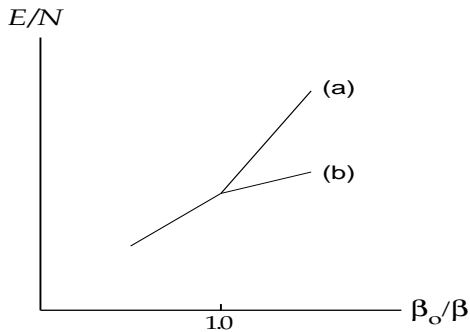}}
\vskip0.5cm
\caption{The energy density as a function of the reduced temperature at the second-order transition. The branch (a) is for $\Delta c >0$, and   (b) is for $\Delta c<0$.}
\label{fig:fig4}  
\end{figure}

One interesting problem to consider is what happens when the positions of the two Gaussian peaks coincide. Since this is the limit where the latent heat $l$ vanishes, one might naively expect that the system would exhibit a second-order transition  of the Ehrenfest classification\cite{hwang}, where  there is a finite discontinuity in specific heat but no latent heat. (Fig.\ref{fig:fig4}).

 However, when we consider the exact zeros of the partition function for the system with two Gaussian peaks, we find there is a branch of zeros other than the one described in ref.\cite{kcl}. For $l \neq 0$, this branch can be neglected, since for generic systems the Gaussian approximation breaks down at this point  due to the contributions from the higher order cumulants. However, for $l=0$ this is no more true and we have to take this branch into account. Because of this, the system exhibits a critical behavior instead of the second-order transition.

\section{Locus and Density of Zeros}
  We consider a canonical partition function of a continuum system, which can be written as in ref.\cite{kcl}
\bea
{\cal M} (t) \equiv Z(\beta)/Z(\beta_0) = \int^\infty_{-\infty} e^{tx} f(x) dx
\eea
where the probability density function is given by
\beq
f(x)= \Omega(x/\beta_0) e^{-x} / \int^\infty_{-\infty} \Omega(x/\beta_0)e^{-x} dx ,
\eeq 
 $t=1-\beta/\beta_0$, $x=\beta_0 E$, $\Omega(E)$ is the density of states at energy $E$ and $\beta_0$ is the inverse of the transition temperature we are interested in. When one is interested in field driven phase transition, one may replace the energy $E$ by the magnetization $M$ and the inverse temperature $\beta$ by the magnetic field $H$ in case of magnetic systems, so on.  We investigate the locus of zeros on the plane of complex temperature $z = r e^{i \theta} \equiv e^t$. 

Now consider the case of interest in this paper, when $f(x)$ is given by summation of two Gaussian peaks up to normalization,
\beq
f(x) = {1 \over \sqrt{2 \pi } \s_1}\exp [-{(x-\m_1)^2 \over 2 \s_1^2}]+ a {1 \over \sqrt{2 \pi } \s_2}\exp [-{(x-\m_2)^2 \over 2 \s_2^2}] 
\label{gau}
\eeq
These two peaks represent two different phases of the system. When $\m_1 \neq \m_2$ the system undergoes the first-order transition. Let us denote two Gaussian functions by $f_1(x)$ and $f_2(x)$. Since we can relabel $f_1$ and $f_2$, and redefine $a$, we may assume $m \equiv  (\mu_2-\mu_1)/2  >0$ without loss of generality. We then have
\bea
{\cal M}(t) &=& \int^\infty_{\infty} e^{tx}[ f_1(x) + a f_2(x)]dx \cr 
          &=& \exp(\psi_1)+ \exp(\psi_2)
\eea
where
\bea
\exp(\psi_1(t)) &\equiv& \int e^{t x} f_1 (x) dx \cr
\exp(\psi_2(t)) &\equiv& \int e^{t x} a f_2 (x) dx. 
\eea 
 From the expressions above, one can easily see that the locations of zeros are given by the solutions to the following equation as in ref.\cite{kcl} 
\beq
\psi_2(t_k)-\psi_1(t_k) = 2 i I_k \equiv i (2k+1) \pi
\eeq 
where $k$ runs through all the integers.
For the double Gaussian distribution the equation above can be rewritten as
\beq
iI_k = {1 \over 2} \ln a + m (t_k) + {\t \sigma^2 \over 2}(t_k)^2
\eeq
where $\tilde \sigma^2 = (\sigma_2^2-\sigma_1^2)/2$. 
This equation is quadratic and easily solved. The solutions are
\bea
t_k^{\pm} &=&  -{m \over \t \s^2} \pm {\sqrt{2} \over |\t \s|}\sqrt{i I_k -{1 \over 2} \ln a + {m^2 \over 2 \t \s^2}} \nonumber \\
&=& -{m \over \t \s^2} \pm {|I_k| \over | \l_k \t \s|} \pm i \ {\rm sign}(I_k) | {\l_k \over \t \s} |
\eea
where 
\beq
\lambda_k \equiv \sqrt{\sqrt{({m^2 \over 2 \tilde \sigma^2}-{\ln a \over 2})^2 + I_k^2 }-({m^2 \over 2 \tilde \sigma^2}-{\ln a \over 2})}
\eeq
 Note that there are two branches of solutions. One passes through the transition point $t=0$ in the thermodynamic limit and the other does not, so the latter was implicitly discarded in ref.\cite{kcl}. As we will see, the second branch closes in toward  $t=0$ as we take the limit $l \to 0$. 

 Now we redefine the variables
\bea
m &=& {N l \over 2} \nonumber \\
I_k &=& {N y_k \over 2} \nonumber \\
\t \s^2 &=& {N \Delta c \over 2} 
\eea 
and consider the thermodynamic limit $N \to \infty$. We then get
\bea
\ln(r_k) &\equiv& \Re (t_k) = -{ m \over \t \s^2} \pm {|I_k| \over |\t \s  \l_k |} \cr
&=&  -{l \over \Delta c} \pm {| y_k | \over \sqrt{\sqrt{({l^2 \over 2 })^2 + y_k^2 (\Delta c)^2}-{l^2 \over 2}} } \nonumber \\
\theta_k &\equiv& \Im(t_k) = \pm {\rm sign}(I_k) |{ \l_k \over \t \s}| \cr
&=& \pm  {\rm sign} (y_k ) \sqrt{\sqrt{{1 \over 4}({l \over \Delta c})^4 + {y_k^2 \over  (\Delta c)^2}}-{l^2 \over 2 (\Delta c)^2}} \label{lo}
\eea
The terms involving $\ln a$ are finite size corrections and vanish in this limit.
We solve the second equation of (\ref{lo}) in terms of $y_k$ to get
\beq
y_k = \pm \theta_k   l  \sqrt{1 + \theta_k^2 ({\Delta c \over l})^2} \label{ang}
\eeq
We substitute (\ref{ang}) into the first equation of (\ref{lo}) to get the locus of zeroes,
\beq
r_\pm = \exp\left[-{l \over \Delta c} \pm {l \over |\Delta c|} \sqrt{ 1 + \theta_k^2 ({\Delta c \over l})^2}\right]
\eeq

We can also obtain the angular density of zeros. By taking formal derivative with respect to the integer $k$, we get 
\bea
\left|{d \theta_k \over d k}\right| &=& {1 \over 2 \sqrt{\sqrt{{1 \over 4}({l \over  \Delta c})^4 + {y_k^2 \over (\Delta c)^2}}-{l^2 \over 2 (\Delta c)^2}}} \nonumber \\
&& \times{2 y_k \pi \over  N (\Delta c)^2 \sqrt{{1 \over 4}({l \over  \Delta c})^4 + {y_k^2 \over  (\Delta c)^2}}} \nonumber \\
&=& {2 \pi l \sqrt{1+\theta^2 ({\Delta c \over l})^2} \over N[l^2 + 2 (\Delta c)^2 \theta^2]}
\eea
Therefore, the angular densities of zeros $g_\pm$ of two branches are given by
\beq
2\pi g_\pm (\theta) \equiv {2 \pi \over N} \left|{dk \over d \theta}\right| =  l {1 + 2 ({\Delta c \over l})^2 \theta^2 \over \sqrt{1 + ({\Delta c \over l})^2 \theta^2 }}
\eeq

\section{First-Order Transitions}

We will now consider both loci of zeros of the partition function  at  first-order transition.  
  Note that all the quantities above depend only on the ratio $l / \Delta c$ except for a overall factor of $l$ in front of $g(\theta)$.\footnote{When both $l$ and $\Delta c$ are zero these quantities are ill-defined and we can no longer use the Gaussian approximation. One then has to take into account higher order cumulants.} When $l / \Delta c  \neq 0$, we get the first-order transition. This is the case considered in ref.\cite{kcl}. There only the locus of zeros near  the transition point $t=0$ were treated carefully since these were the only things of interest. In fact, for generic systems we expect that the Gaussian approximation breaks down away from the transition point $t=0$ due to the higher-order cumulants. 

\begin{figure}
\epsfxsize=8.00cm
\epsfysize=11.5cm
\vskip0.1cm
\centerline{
\epsfbox{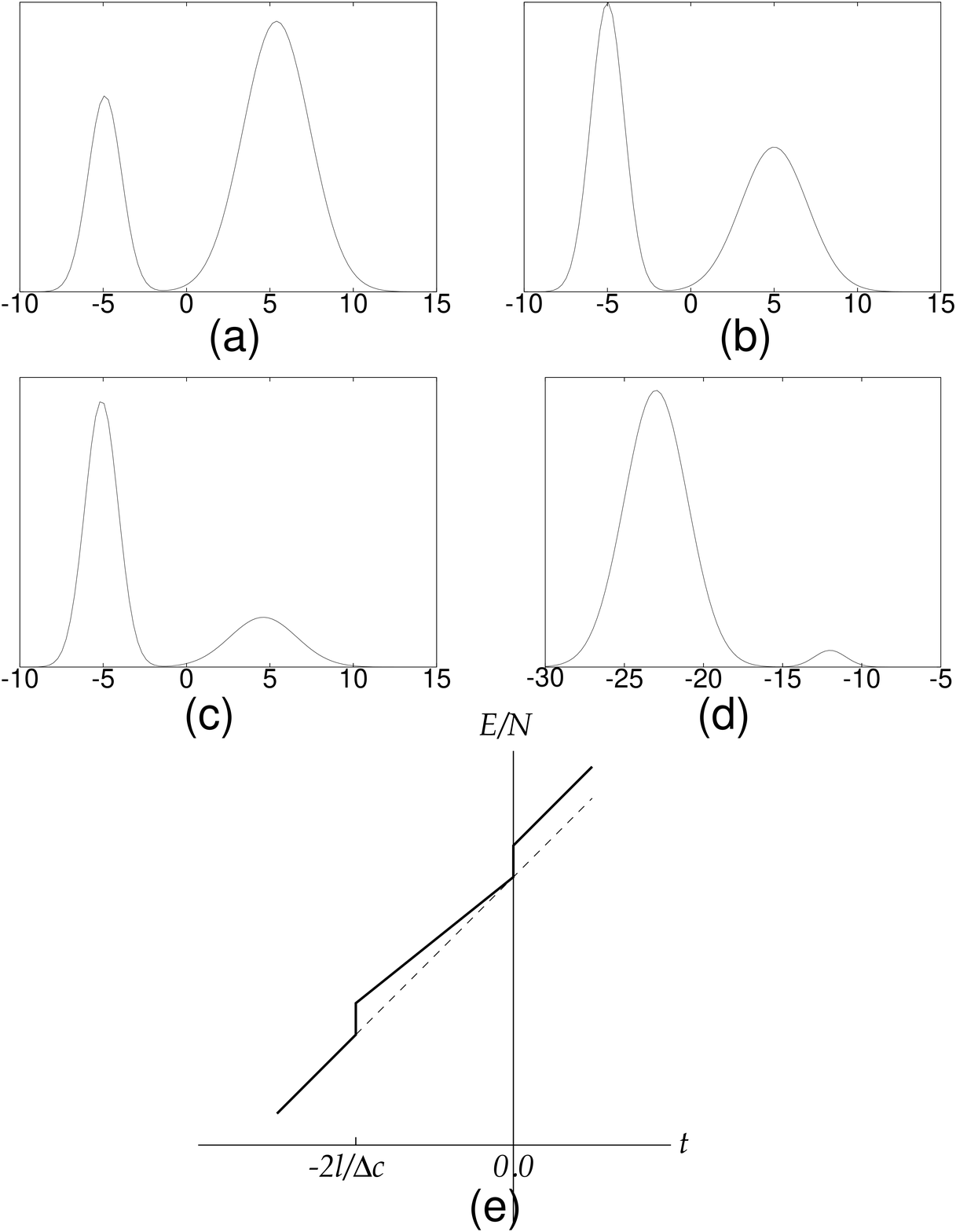}}
\vskip0.5cm
\caption{Qualitive behavior of $e^{tx}\Omega(x)$ versus x, where we take $\mu_{1,2}=\pm 5$ in this example. (a) For $t >0$ the peak 2 dominates (b) At $t=0$ the weight of the two Gaussian peak is equal by construction, meaning the areas under the curves are same. 
(c) As $t$ decreases below $0$, the weight of peak 1 becomes larger. The first-order transition has occurred with the latent heat $l$ and the discontinuity in the specific heat $\Delta c$. The positions of the peaks begin to get shifted to th left, with peak 2 moving faster. 
(d) For $t<-2l/\Delta c$, the weight of the peak 2 becomes larger than that of 1 again.  At this point the peak 2 is at the left of the peak 1, so this is another first-order transition, with the latent heat $l$ and the discontinuity in the specific heat $-\Delta c$. (e) Schematic diagram of energy versus reduced temperature for the system with double Gaussian peaks. Note that there are two first-order transition with the same latent heat but opposite sign for the discontinuity in the specific heat. The transition at $t=-2l/\Delta c$ is discarded for generic systems.}
\label{gau1}  
\end{figure}

Let us elaborate on this point.  The locus of zeros cross the real axis at $t=0$ and $t=-2l/\Delta c$, indicating there are two phase transitions. This can be easily understood. The probability density  at arbitrary temperature is given by
\bea
e^{tx} \Omega(x) &=& {e^{tx} e^{-(x-x_1)^2/(2 \s_1^2)} \over \sqrt{2 \pi \sigma_1}}+{e^{tx} e^{-(x-x_2)^2/(2 \s_2^2)}\over \sqrt{2 \pi \sigma_2}}  \nonumber \\
&=& {e^{-(x-x_1-\s_1^2 t)^2/(2 \s_1^2)+x_1 t + \s_1^2 t^2/2} \over \sqrt{2 \pi \sigma_1}} + {e^{-(x-x_2-\s_2^2 t)^2/(2 \s_2^2)+x_2 t + \s_2^2 t^2/2} \over \sqrt{2 \pi \sigma_2}}
\eea
We see that for nonzero $t$ the positions of the peaks are shifted, and also the relative weights change. We see that the position of the peak for large $\s_i$ gets  shifted by a larger amount for given temperature change, consistent with the fact that it has larger specific heat.  The  weight of the peak 1 relative to the peak 2 is given by:
 \beq
  w1/w2 \equiv \exp ( {(\s_1^2-\s_2^2)\over 2} t^2 + (x_1-x_2) t) 
\eeq
 By construction, at $t=0$, the weight of two Gaussian peak is equal. Assuming $\s_2 > \s_1$, we see that for $t>0$ the peak labeled by 2 dominates. When $t$ becomes slightly negative, then the peak 1 dominates. Also the positions of the Gaussian peaks get shifted to left, but the peak 2 moves faster.  For $t< -(x_1-x_2)/(\s_1^2-\s_2^2) $ the peak 2 goes to the left of the peak 1. At $t=-2l/\Delta c $, the weight of the peak 2  become equal to that of 1 again, and the peak 2 is dominant for $t< -2l/\Delta c$.   Therefore at this temperature there is another first-order transition with latent heat $l$ and specific heat change $-\Delta c$. We can make similar arguments for $\s_1 > \s_2$. This process is depicted in Fig.\ref{gau1}. 
 
This mechanism works only if we trust that the Gaussian form given in (\ref{gau}) is exact. However, for a generic system, this is just a leading   truncation of the cumulant expansion 
\beq
 \exp[-Nf(x)]= \exp[-N(f(x_0)+{f''(x_0) \over 2} (\Delta x)^2 +{f''(x_0) \over 3!} (\Delta x)^3 + \cdots)], 
\eeq
so the higher order cumulants can be ignored only when $\Delta x << O(1/\sqrt{N})$. But at the first-order transition at $t=-2l/\Delta c$, the system is dominated by the peaks which are located at the distances of $O(1)$ from the positions of the peaks at $t=0$. Therefore the higher-order cumulants would contribute, and we cannot trust the picture above. However, when  $\Delta x << O(1/\sqrt{N})$, or when the higher order cumulants are extremely small due to some reason,  transition at $t= -2l/\Delta c$ cannot be neglected anymore. In particular,  in the limit $l \to 0$, the second branch touches the first branch, preventing the system from exhibiting the second-order transition.

\vskip1.0cm

  The behaviors of the loci of zeros for various values of $\Delta c/l$ are depicted in Fig.1,2 and 3 in complex $z \equiv \exp(t)$ plane.
 $t_+$ is the outer curve and $t_-$ is the inner curve. Only the zeros in the first Riemann sheet are shown. When $\Delta c/l>0$ $(<0)$, $t_+$ $(t_-)$ passes through $t=0$, and becomes unit circle as one approaches the symmetric limit, $\Delta c/l \to 0$.  This is consistent with  Lee-Yang's unit circle theorem. The other branch $t_-$ $(t_+)$ degenerates to origin.(goes to infinity.)  The loci intersect the real axis orthogonally as long as $l \neq 0$.

\begin{figure}
\epsfxsize=8.00cm
\epsfysize=11.82cm
\vskip0.1cm
\centerline{
\epsfbox{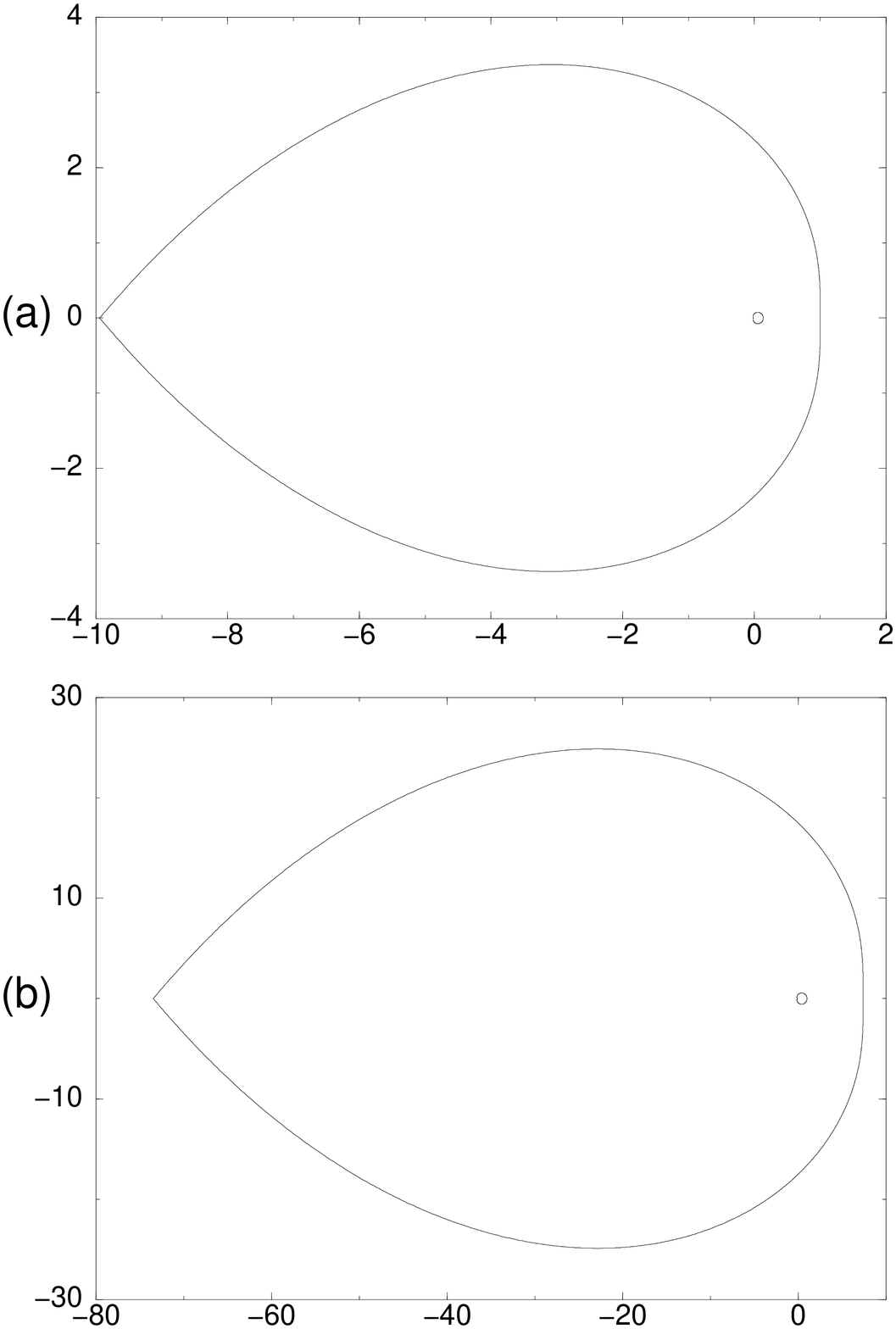}}
\vskip0.5cm
\caption{(a) $\Delta c/l = 1$. As $\Delta c/l \to 0$, the outer curve becomes a unit circle and the inner curve degenerates to origin.\\ (b) $\Delta c/l = -1$. As $\Delta c/l \to 0$, the inner curve becomes a unit circle and the outer curve goes away to infinity.}
\label{fig:fig1}  
\end{figure}

\begin{figure}
\epsfxsize=8.00cm
\epsfysize=11.82cm
\vskip0.1cm
\centerline{
\epsfbox{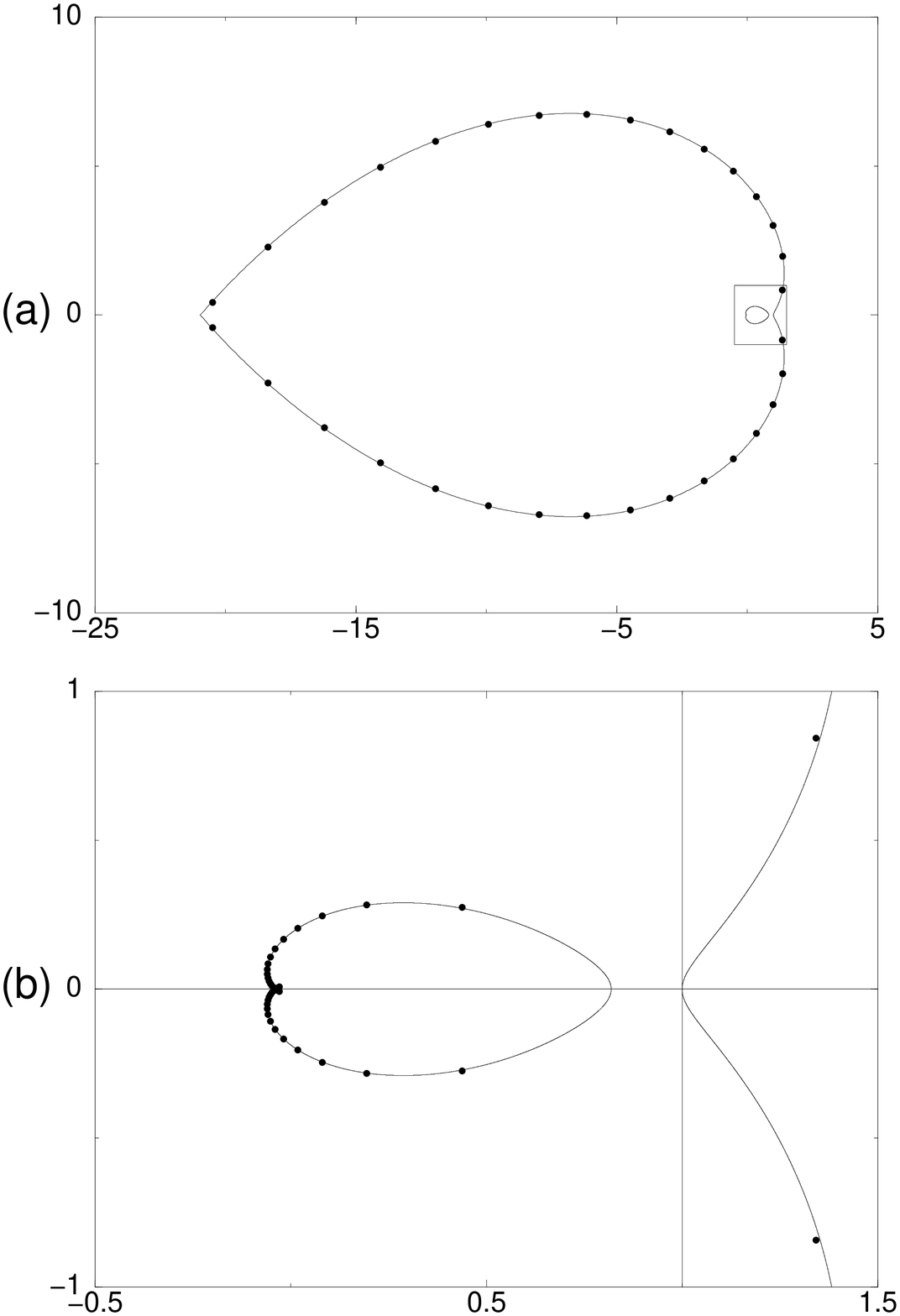}}
\vskip0.5cm
\caption{(a) $\t \sigma^2/m=\Delta c/l = 10$. 
The zeros for finite $N$ are also plotted as dots,
 for $2 \t \sigma^2 = N \Delta c=10$ and $a=1.0$. 
The dots deviate from the curve if $a \neq 1.0$.\\
 (b)  Magnification of the box in (a). The horizontal line indicates the real axis, and the vertical line is given by  $Re(z)=1$. Note that the outer curve passes through the point $z=1.0$, and the angles between both loci and the real axis are 90 degrees.}
\label{fig:fig2}  
\end{figure}

\begin{figure}
\epsfxsize=8.00cm
\epsfysize=11.82cm
\vskip0.1cm
\centerline{
\epsfbox{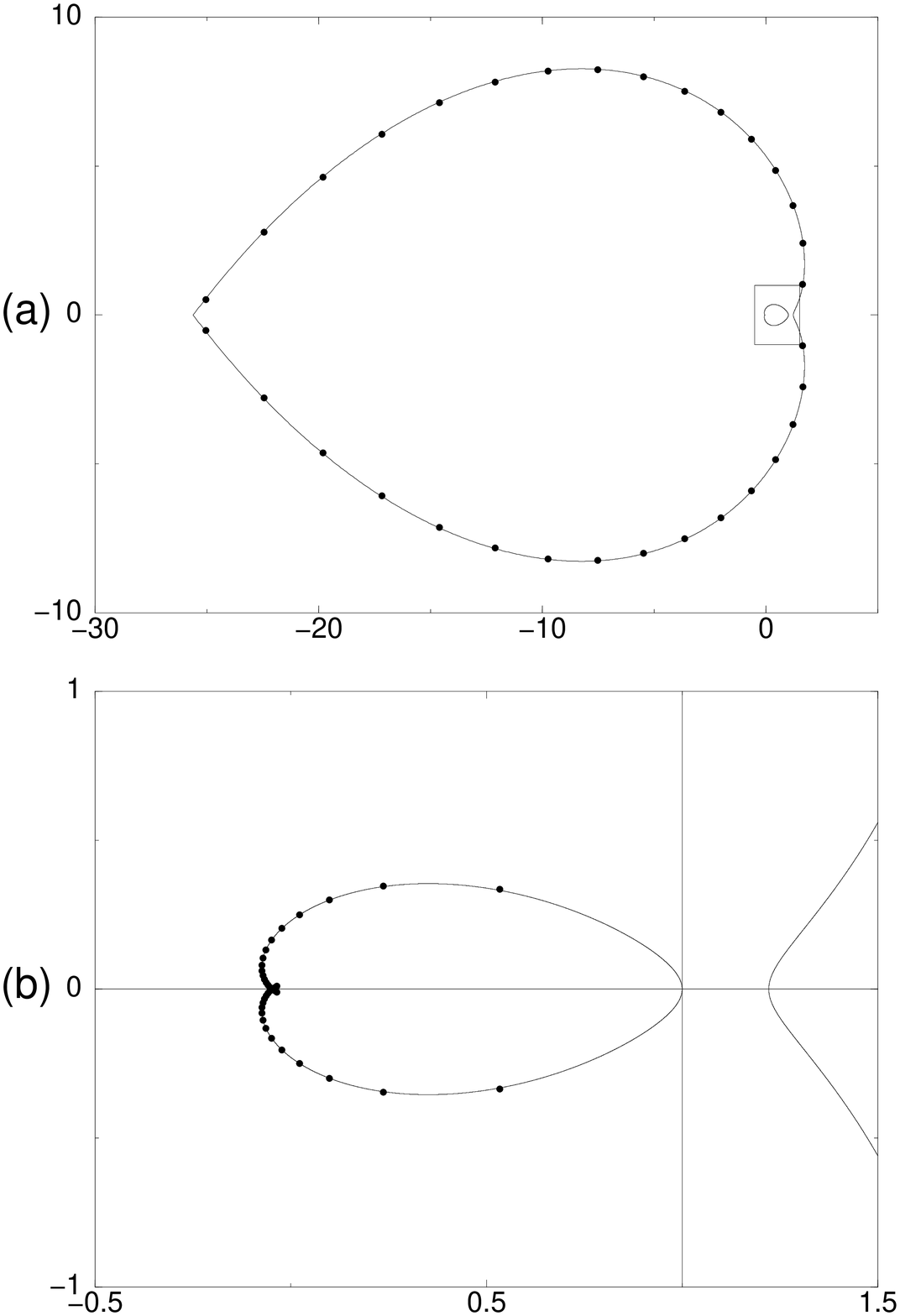}}
\vskip0.5cm
\caption{(a)$\Delta c/l = -10$. The dots indicate zeros for $2 \t \sigma^2 = N \Delta c=-10$ and $a=1.0$. \\
(b) Magnification of the box in (a). It is the inner curve which passes through $z=1.0$ in this case.}
\label{fig:fig3}  
\end{figure}

\section{$l \to 0$ limit and the critical behavior}

\begin{figure}
\epsfxsize=8.00cm
\epsfysize=11.82cm
\vskip0.1cm
\centerline{
\epsfbox{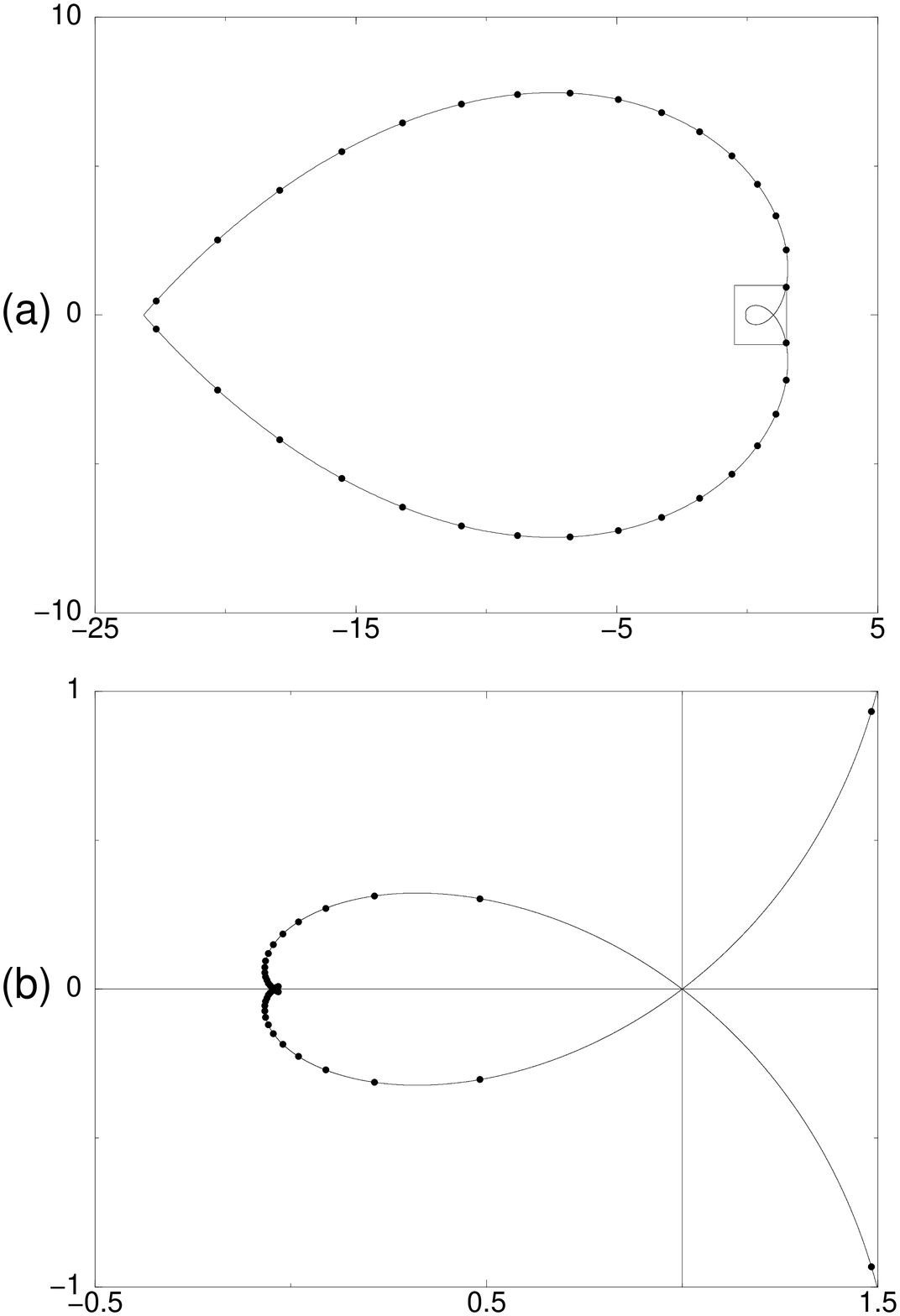}}
\vskip0.5cm
\caption{(a) $l=0$. The zeros for $N|\Delta c|=10$ are plotted, with the same notation as before.\\
 (b) Magnification of the box in (a).  Note that the two loci are now joined to form a single curve, and intersects the real axis at $z=1.0$ with the angle of 45 degrees. Both $\Delta c>0$ (Fig.\ref{fig:fig2}) and $\Delta c<0$ (Fig.\ref{fig:fig3}) approach this limit as $l \to 0$.}
\label{fig:fig5}  
\end{figure}

The limit $l/\Delta c=0$  may be considered as  the opposite limit from the symmetric case $\Delta c/l = 0$.   Now the two loci $t_\pm$ which were separate when $l \neq 0$, touch each other at $\theta=0$ and form a single curve. (Fig.\ref{fig:fig5}). Their loci are given by
\beq
r_\pm = \exp(\pm |\theta|)
\eeq
The density of zeros is
\beq
2 \pi g(\theta) = 2 \pi (g_+(\theta) + g_-(\theta)) = 4 \Delta c |\theta|.
\eeq

Note that $g(\theta)$ is zero at $\theta=0$, consistent with the fact that the first derivative of the partition function has no discontinuity.  The loci intersect the real axis with the angles of 45 degrees, and the intermediate region dominated by peak 1 with the smaller specific heat,  which used to separate two domains dominated by peak 2 with the larger specific heat, touches the  real axis at just one point. Therefore the system is dominated by the peak 2 for $t \neq 0$,  and the peak 1 has same weight as the peak 2 only at $t=0$, when their positions coincide. The qualitative behaviors of the two peaks for $t>0$ and $t<0$ are same as the ones depicted in Fig.2(a) and Fig.2(d). 

Therefore the system is exhibiting a critical behavior where it is just on the verge of making a phase transition. However, in contrast to many familiar examples of critical behavior, the specific heat near $t=0$ remains finite instead of blowing up. At this stage it is not yet clear whether there is an example of a discrete system whose critical behavior at the thermodynamic limit can be described by this model.

\acknowledgments

This work was supported in part by the Ministry of Education, Republic of 
Korea
through a grant to the Research Institute for Basic Sciences, Seoul National
University, in part by the Korea Science Foundation through Research Grant to
the Center for Theoretical Physics, Seoul National University, and in part by the Japanese Society for Promotion of Science through the Institute of Physics, University of Tokyo.


\begin{thebibliography}{99}
\bibitem{yang72}
C.N.Yang, in {\it Phase Transitions and Critical Phenomena Vol.  1}
edited by C.  Domb and M.S.  Green Academic Press (1972)
\bibitem{ly} C. N. Yang and T. D. Lee, Phys. Rev. {\bf 87}, 404 (1952); T. D. Lee and C. N. Yang, {\it ibid.}  {\bf 87}, 410 (1952).

\bibitem{fisher65}  
M.  E.  Fisher, in {\it Lectures in Theoretical Physics} (University of
Colorado Press, Boulder, 1965), Vol.  7c

\bibitem{ono68} 
S. Ono, Y. Karaki, M. Suzuki and C. Kawabata, J. Phys. Soc. Japan
{\bf 25}, 54 (1968)

\bibitem{pearson82} 
R.B.  Pearson, Phys. Rev. {\bf B26}, 6285 (1982)

\bibitem{abe67} 
R.  Abe, Prog.  Theor.  Phys.  {\bf 38}, 322, (1967)

\bibitem{itzykson83} 
C.Itzykson, R.B.  Pearson, and J.B.  Zuber, Nucl.  Phys.  {\bf B220},
415 (1983)

\bibitem{privman87} 
M.L.  Glasser, V.Privman, and L.S.  Schulman, Phys. Rev.  {\bf B35},
1841 (1987)

\bibitem{bhanot87} 
G.  Bhanot, R.  Salvador, S.  Black, P.  Carter and R.  Toral, Phys
Rev.  Lett.  {\bf 59}, 803 (1987)

\bibitem{kcl93} Koo-Chul Lee,  Phys. Rev. E {\bf 48}  3459 (1993). 


\bibitem{kcl94} Koo-Chul Lee,  Phys. Rev. Lett.  {\bf 73}, 2801 (1994)

\bibitem{kcl99} Sooyeul Lee, Seunghwan Kim, Seon Hee Park, Hyeon-Bong Pyo and Koo-Chul Lee,  Phys. Rev. B {\bf 60}  9256 (1999). 

\bibitem{kcl} Koo-Chul Lee,  Phys. Rev. E {\bf 53}  6558 (1996). 

\bibitem{hwang}P. Ehrefest, Proc. Roy. Acad. Amsterdam, {\bf 36}, 153 (1933); {\it Leiden Commun.}, Suppl. 75b(1933);K. Huang, {\it Statistical Mechanics} 2nd ed., (Wiley, New York, 1987),  p. 35.
 

 \end{thebibliography}
\end{document}